\documentclass[12pt]{iopart}

\usepackage{latexsym}
\usepackage{graphicx}
\usepackage{iopams}  

\begin{document}

\title[Anomalous chain dynamics in polymer blends]
{Static and dynamic contributions to anomalous chain dynamics in polymer blends}

\author{Marco Bernabei$^{1}$, Angel J. Moreno$^{2}$ \footnote[3]{To whom correspondence should be 
addressed (wabmosea@ehu.es)}, J. Colmenero$^{1,2,3}$}

\address{$^{1}$ Donostia International Physics Center, Paseo Manuel de Lardizabal 4,
E-20018 San Sebasti\'{a}n, Spain.}

\address{$^{2}$ Centro de F\'{\i}sica de Materiales (CSIC, UPV/EHU) and Materials Physics Center MPC, 
Paseo Manuel de Lardizabal 5, E-20018 San Sebasti\'{a}n, Spain.}

\address{$^{3}$ Departamento de F\'{\i}sica de Materiales, Universidad del Pa\'{\i}s Vasco (UPV/EHU),
Apartado 1072, E-20080 San Sebasti\'{a}n, Spain.}

\begin{abstract}

By means of computer simulations, we investigate the relaxation of the Rouse modes in a simple bead-spring model
for non-entangled polymer blends.  Two different models are used for the fast component, namely
fully-flexible and semiflexible chains. The latter are semiflexible
in the meaning that static intrachain correlations are strongly non-gaussian 
at all length scales. The dynamic asymmetry in the blend is strongly enhanced by decreasing temperature,
inducing confinement effects on the fast component. The dynamics of the Rouse modes show very different trends
for the two models of the fast component. For the fully-flexible case,  the relaxation times
exhibit a progressive deviation from Rouse scaling on increasing the dynamic asymmetry. This anomalous effect has
a dynamic origin.  It is not related to particular static features of the Rouse modes, which indeed
are identical to those of the fully-flexible homopolymer, and are not modified by the dynamic asymmetry in the blend. 
On the contrary, in the semiflexible case the relaxation times exhibit approximately
the same scaling behaviour as the amplitudes of the modes. This suggests that the origin of
the anomalous dynamic scaling for semiflexible chains confined in the blend is esentially of static nature.
We discuss implications of these observations for the applicability of theoretical approaches to chain dynamics
in polymer blends.

\end{abstract}


\maketitle

\section{Introduction}

Miscible polymer blends exhibit dynamic disparity. By starting from two homopolymers with different mobilities,
two separated segmental relaxations are still found in the blend state \cite{blendsrev,maranas}.
When the two homopolymers exhibit very different glass transition temperatures, and the concentration
of the fast component is low, the respective time scales in the blend 
can differ by orders of magnitude, leading to strong dynamic asymmetry \cite{lorthioir}.
This effect is enhanced on decreasing the temperature, and can be of even 12 decades 
for high dilution of poly(ethylene oxide) (PEO) in
poly(methyl methacrylate) (PMMA) \cite{lutz}. In conditions of strong dynamic asymmetry, the motion of the
fast component is confined by the slowly relaxing matrix formed by the slow component.
Unusual dynamic features arise for the fast component in this case. Neutron scattering experiments 
in the blend PEO/PMMA reveal decoupling between self-motions and intrachain collective relaxation
for the PEO \cite{niedzwiedz}. Atomistic simulations of the former system
reveal logarithmic decays of the scattering functions probing  the segmental relaxation
of the PEO \cite{genix}. Moreover, the Rouse modes of the PEO chains exhibit 
anomalous behaviour \cite{brodeck} (see below).
All these observations have also been found in simulations of a simple blend of bead-spring chains
with strong dynamic asymmetry \cite{blendsjcp,blendsjphys,prlrouse},
suggesting that they are generic features of real blends.

The simple bead-spring chains of the blends investigated in Refs.~\cite{blendsjcp,blendsjphys,prlrouse}
are fully-flexible. No intramolecular barriers are implemented. 
The chain length, $N$, in all the investigated cases is smaller than the entanglement value.
In such conditions, the simulations for the homopolymer system are 
consistent with expectations from the Rouse model \cite{doibook,kremer,bennemannrouse,shaffer,kreer}.
This is also the case for the slow component in the blend state \cite{prlrouse}. However a rather different scenario
is found for the fast component in the blend \cite{prlrouse},
which only shows Rouse dynamics for vanishing dynamic asymmetry. 
Increasing the latter (by decreasing temperature) induces a progressive deviation from Rouse-like behaviour 
for {\it dynamic} observables of the fast component. These include intrinsic non-exponentiality
of the Rouse modes and anomalous scaling of the corresponding relaxation times, 
$\tau_p \sim p^{-x}$, where $p$ is the mode index. On increasing the dynamic asymmetry
the exponent increases from the Rouse prediction, $x = 2$, 
to  values $x \lesssim 3.5$.
The origin of these features is dynamic. 
Indeed the static amplitudes of the Rouse modes are not affected by blending, 
and show  gaussian, Rouse-like, scaling as in the homopolymer state \cite{prlrouse}.

The mentioned dynamic crossover, on increasing the dynamic asymmetry, 
from $\tau_p \sim p^{-2}$ to $\tau_p \sim p^{-3.5}$ in the {\it non-entangled} fast component
is strikingly similar to that observed for entangled homopolymers 
on increasing the mode wavelength, $N/p$,  beyond the entanglement length \cite{shaffer,kremerprl,padding}.
The crossover for the fast component in the blend is observed even in the limit of short chains. 
Thus, it was concluded that this feature is entirely controlled by the dynamic asymmetry in the blend,
and not by a characteristic length scale \cite{prlrouse}. 
Related observations have been presented by simulations of short fully-flexible chains 
in  matrixes of fixed obstacles \cite{yamakov1,yamakov2,changprl,changjcp}. 
Rouse modes were not analyzed
but a crossover, similar to the observation in homopolymers on increasing $N$, was found.
Thus, the diffusivities changed from $D \sim N^{-1}$ to $D \sim N^{-2}$
on increasing the concentration of obstacles. Concomitantly, the end-to-end relaxation time 
changed from $\tau_{\rm e} \sim N^2$ to $\tau_{\rm e} \sim N^3$. Again, these observations
are neither related to static intramolecular features nor to characteristic length scales,
being entirely controlled by the concentration of obstacles \cite{changprl,changjcp}.

In summary, the former results of 
Refs.~\cite{prlrouse,yamakov1,yamakov2,changprl,changjcp} reveal a crossover
to entangled-like dynamic features in non-entangled chains. This crossover is 
not connected to particular static features of the intrachain correlations.
It has a entirely dynamic origin, related to the strength of the confinement 
effects induced by the surrounding matrix. Theories based on generalized Langevin equations (GLE) 
introduce a memory kernel accounting for the slow relaxation of density fluctuations
around the tagged chain \cite{schweizer}. The Rouse model, which neglects memory effects,
arises from such theories
in the limit of fast relaxation of the kernel \cite{schweizer,kimmich}.
Memory effects are enhanced on increasing the chain length beyond the entanglement
value, and the theory reproduces non-exponentiality and anomalous scaling of the Rouse modes 
in entangled homopolymers \cite{kimmich}. It has been suggested that GLE methods
may also account for the analogous dynamic features exhibited by the non-entangled fast component
in polymer blends \cite{prlrouse}, through the incorporation in the kernel
of the memory effects induced by the matrix, formed by the slow component in the blend 
or by the fixed obstacles in the systems of Refs.~\cite{yamakov1,yamakov2,changprl,changjcp}. 

As mentioned above, the simulations 
of Refs.~\cite{prlrouse,yamakov1,yamakov2,changprl,changjcp} were performed
for fully-flexible chains. In this article we briefly review the results of \cite{prlrouse}
and present new results for blends
where the fast component is semiflexible, i.e., it has intramolecular barriers. 
These are introduced by implementing bending and torsion potentials.
It is well-known that the presence of strong intramolecular barriers violates
the Rouse assumption of gaussian behaviour  
for the static and dynamic intrachain correlations. Indeed significant non-Rouse effects 
appear as chains become stiffer \cite{paul,krushev,bulacu1,steinhauser,steinhauserrev,bernabei1}.
In the same spirit as the Rouse model for fully-flexible chains, 
phenomenological models for semiflexible polymers usually 
model the interactions of the tagged chain with the surroundings
by means of a friction term and random forces \cite{allegra,harnau1}.
Thus, memory effects induced by slow density fluctuations of the matrix are neglected,
and non-Rouse effects are of intramolecular origin. The latter is entirely related
to static contributions and the amplitudes and relaxation times of the Rouse modes
follow the same scaling behaviour \cite{steinhauserrev}.
 
The chains studied in this work are semiflexible
in the meaning that they can be deformed but, unlike fully-flexible chains, 
the static intramolecular correlations are strongly non-gaussian within all the chain
length scale.  We investigate the effect of blending on
the dynamic scenario observed for the semiflexible homopolymer,
aiming to discriminate static and dynamic contributions 
to anomalous chain dynamics in the blend. We find strong deviations from Rouse behaviour.
However, as in the homopolymer state, 
the amplitudes and relaxation times of the Rouse modes follow 
almost the same scaling behaviour, i.e, in this case the origin of the anomalous, non-Rouse, features found for chain
dynamics in the blend is  esentially  static. We conclude that memory effects induced by the
slow matrix are not significant for semiflexible chains in blends,
at least for the range of dynamic asymmetry investigated here, and when
the chains are non-gaussian over all their length scale. Presumably,
memory effects will become relevant in semiflexible chains which are long enough, at large length scales for which 
intramolecular static correlations recover gaussian statistics.

The paper is organized as follows. We describe the model and give simulation details in Section 2.
Results for static and dynamic observables are presented in Section 3 and discussed in Section 4. 
Conclusions are given in Section 5.

\section{Model and simulation details}

We simulate a binary mixture of bead-spring chains. Monomers within a same chain are identical, 
i.e, of the same species (A or B). Each chain consists of $N = 21$ monomers, which is below
the entanglement lenght \cite{putz}. 
All monomers have identical mass $m = 1$. Non-bonded interactions between monomers are given
by the potential
\begin{equation}
V_{\alpha\beta}(r) = 4\epsilon[(\sigma_{\alpha\beta}/r)^{12} - 7c^{-12} + 6c^{-14}(r/\sigma_{\alpha\beta})^{2}],
\end{equation}
with $\epsilon=1$, $c = 1.15$ and $\alpha$, $\beta$ $\in$ \{A, B\}.
Potential and forces are continuous at the cutoff $r _{\rm c} = c\sigma_{\alpha\beta}$.
With this cutoff the potential $V(r)$ has no local minima and is purely repulsive.
The interaction diameters are $\sigma_{\rm AA} =1.6$, $\sigma_{\rm AB} = 1.3$, 
and $\sigma_{\rm BB}=1$.
Chain connectivity is introduced by a FENE potential \cite{grest}, 
\begin{equation}
V^{\rm FENE}_{\alpha\alpha}(r) = -kR_0^2 \epsilon\ln[ 1-(R_0\sigma_{\alpha\alpha})^{-2}r^2 ],
\end{equation}
between consecutive monomers, with $k=15$ and $R_0 = 1.5$. 
Intramolecular barriers are implemented by means of the  bending and torsional potentials
proposed in Refs.~\cite{bulacu1,bulacu2}.
The bending potential $V_{\rm B}$ is defined as 
\begin{equation}
V_{\rm B}(\theta_i) = (\epsilon K_{\rm B}/2 )(\cos\theta_i - \cos\theta_0)^2,
\label{eq:potben}
\end{equation}
where $\theta_i$ is the bending angle between consecutive monomers $i-1$, $i$ and $i+1$.
We use $\theta_0 = 109.5^{\rm o}$  for the equilibrium bending angle \cite{bulacu1,bulacu2}.
The torsional  potential $V_{\rm T}$ is defined as
\begin{equation}
V_{\rm T}(\theta_{i},\theta_{i+1},\phi_{i,i+1}) = 
\epsilon K_{\rm T}\sin^3 \theta_{i} \sin^3 \theta_{i+1} \sum_{n=0}^3 a_n \cos^n \phi_{i,i+1},
\label{eq:pottor}
\end{equation}
where $\phi_{i,i+1}$ is the dihedral angle between the two planes defined by the sets of monomers
($i-1$, $i$, $i+1$) and ($i$, $i+1$, $i+2$). Following Refs.~\cite{bulacu1,bulacu2}, we use 
the values $a_0=3.00$, $a_1=-5.90$, $a_2=2.06$, and $a_3=10.95$.
In all the investigated systems the A-chains are fully-flexible, i.e., $K_{\rm B} = K_{\rm T} =0$.
We investigate two models (I and II) for the B-chains. In the model I all the B-chains are fully-flexible.
In the model II all the B-chains are semiflexible, with $K_{\rm B} =15$ and $K_{\rm T} =0.5$.

In the following, temperature $T$, time $t$ and distance
are given respectively in units of $\epsilon/k_B$ (with $k_B$ the Boltzmann constant), 
$\sigma_{\rm BB}(m/\epsilon)^{1/2}$ and $\sigma_{\rm BB}$.
The blend composition is  $x_{\rm B} = N_{\rm B}/(N_{\rm A} + N_{\rm B})$,
with $N_{\alpha}$ the total number of $\alpha$-monomers in the system. 
All simulations are performed at fixed composition $x_{\rm B} = 0.3$.
Most of the simulated systems have 105 A-chains and 45 B-chains. 
At the lowest  investigated temperatures
we have used smaller systems of 49 A-chains and 21 B-chains.
We have performed additional simulations for B-homopolymers of $N =21$,
with a system size ranging from 200 to 500 chains  according to the simulated temperature.
All the simulated systems have a packing fraction 
$\phi = [\pi/(6V)](N_{\rm A}\sigma^3_{\rm AA}+N_{\rm B}\sigma^3_{\rm BB}) = 0.53$, with $V$
the volume of the cubic simulation box. We implement periodic boundary conditions. 
Equations of motion are integrated in the velocity Verlet scheme \cite{frenkel},
with a time step ranging from $10^{-4}$ to $4 \times 10^{-3}$ according to the simulated temperature.
After equilibration at each state point, the corresponding production run 
is performed in the microcanonical ensemble. The longest production runs are of about 400 million
time steps. Averages are performed over up to four independent boxes,
with 20 equispaced time origins per simulated box.

\section{Results and discussion}

\subsection{Chain size and mean squared displacements}

\begin{figure}
\begin{center}
\includegraphics[width=0.50\textwidth]{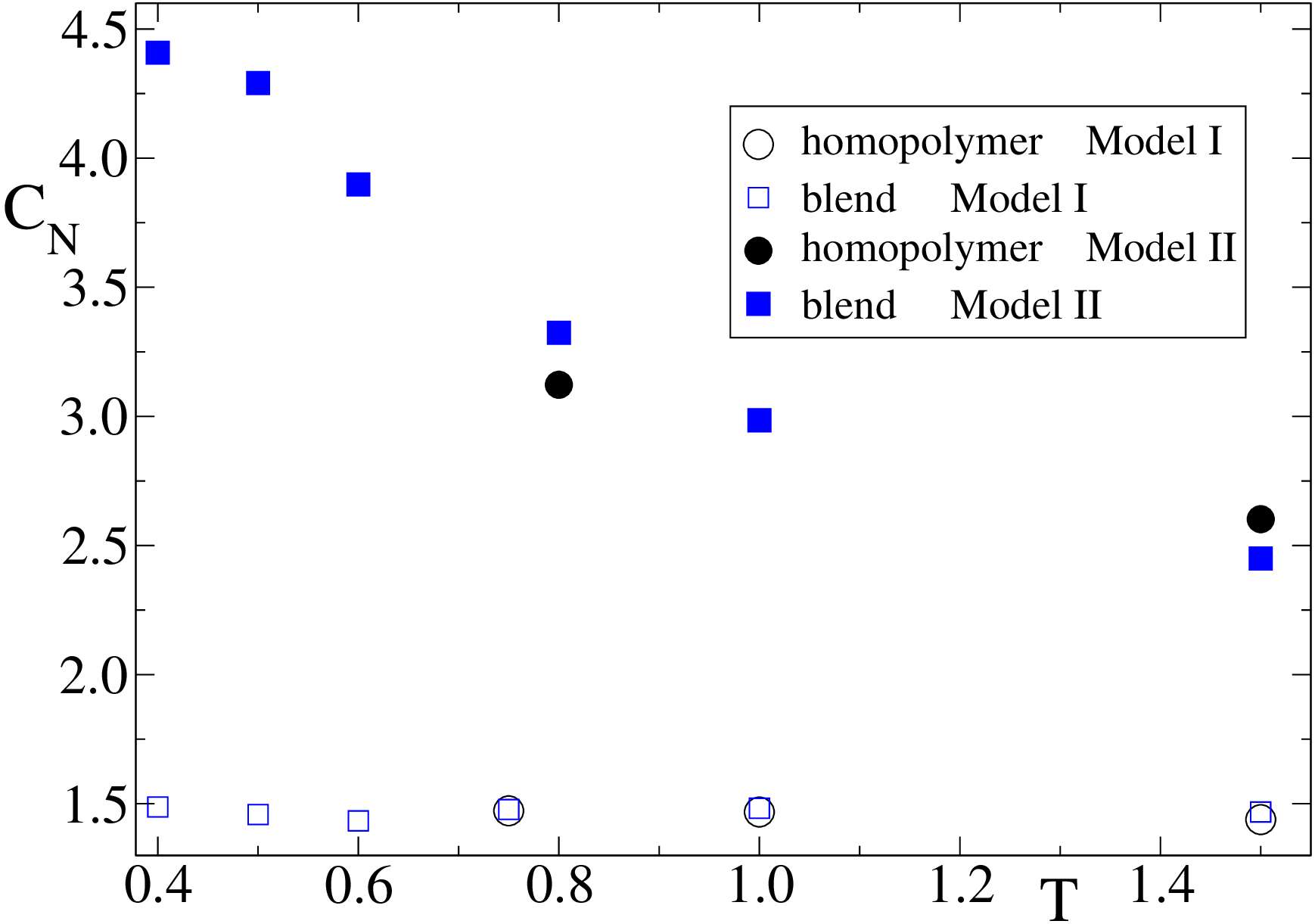}
\end{center}
\caption{Temperature dependence of the characteristic ratio of the fully-flexible 
and semiflexible B-chains, both in the homopolymer and in the blend.}
\label{fig:cinfty}
\end{figure}

The selected values of the bending and torsion constants $(K_{\rm B}, K_{\rm T})$
considerably stretch the semiflexible B-chains in comparison with the fully-flexible case.
This can be quantified by the characteristic ratio,
$C_N = \langle R_{\rm e}^2 \rangle/(N\langle b^2 \rangle)$, where
$\langle R_{\rm e}^2 \rangle$ and $\langle b^2 \rangle$ are respectively the
average squared end-to-end radius and bond length of the B-chains.
Figure~\ref{fig:cinfty} shows results of $C_N$ for the fully-flexible and semiflexible
B-chains. The fully-flexible B-chains exhibit an almost $T$-independent value $C_N \lesssim 1.5$,
both in the homopolymer and in the blend. On the contrary, for the semiflexible
B-chains decreasing temperature yields effectively higher intramolecular barriers.
Thus, the chains become stiffer, and $C_N$ shows a strong increase 
on decreasing temperature. In the blend we find a variation 
of an 80 \%, from $C_N= 2.45$ to $C_N = 4.40$,
over the investigated $T$-range.

\begin{figure}
\begin{center}
\includegraphics[width=0.83\textwidth]{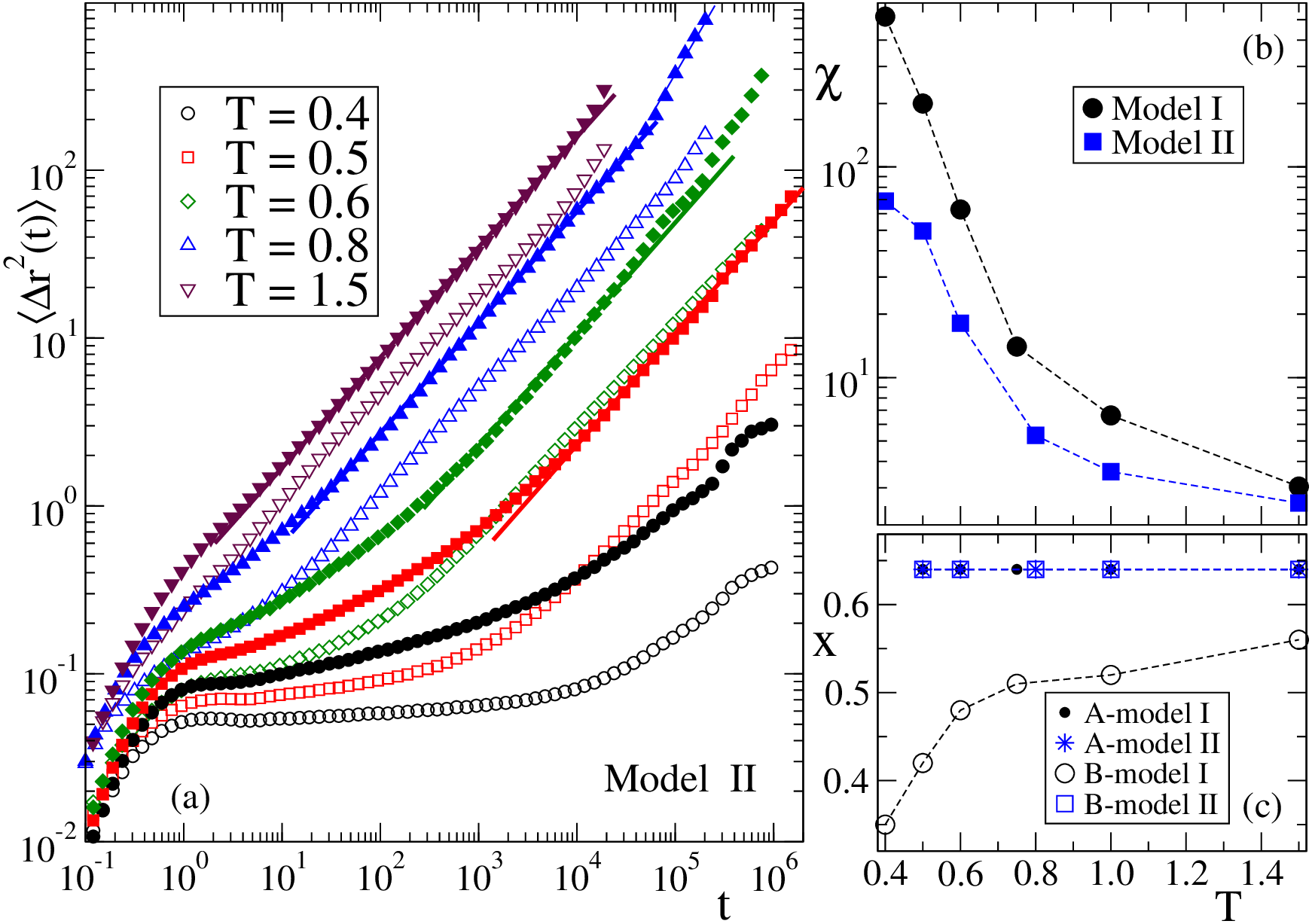}
\end{center}
\caption{Panel (a): MSD in the model II at the different investigated temperatures.
Empty and filled symbols correspond respectively to A- and B-monomers. Data sets
with a same symbol code correspond to a same temperature. Thick straight lines indicate
approximate power law behaviour $\langle \Delta r^2 (t)\rangle \sim t^x$ with $x = 0.64$.
The thin line corresponds to the diffusive limit $\langle \Delta r^2 (t)\rangle \propto t$.
Panel (b): temperature dependence of the dynamic asymmetry (see text)
for both models I and II. Panel (c): temperature dependence of the exponent $x$
of the observed power law in the MSD (see above), for both models I and II.}
\label{fig:msdasym}
\end{figure}

Figure~\ref{fig:msdasym}a shows results for the mean squared displacement (MSD) of both
species A and B in the model II, at all the investigated temperatures. 
After the initial ballistic regime, a plateau arises for both components at $T < 1.5$,
and extends over longer time scales on decreasing temperature. 
This reflects the usual caging regime observed in glass-forming systems
when approaching a glass transition. In analogy with previous
observations for the model I \cite{blendsjcp,blendsjphys}, 
there is a progressive separation, on decreasing $T$, of the
time scales of the A- and B-monomers. We quantify the dynamic asymmetry in the blend
as the ratio $\chi = \tau_{\rm A}/\tau_{\rm B}$, where $\tau_{\alpha}$ is the time
at which the MSD reaches the value $\langle \Delta r^2_{\alpha}(\tau_{\alpha}) \rangle = 0.45$.
This roughly corresponds to the time scale of the structural $\alpha$-relaxation (segmental relaxation).
Figure~\ref{fig:msdasym}b shows the $T$-dependence of the dynamic asymmetry, both for models I and II.
In both cases the dynamic asymmetry becomes stronger on decreasing $T$. 
However, this effect is less pronounced
in the model II, which shows a lower $\chi$ than the model I at the same temperature, 
blend composition and packing fraction.
This is not surprising since the only difference between both models is the strength of the intramolecular
barriers in the B-chains. Thus, dynamics of the B-monomers in the semiflexible homopolymer 
are strongly slowed down respect to the fully-flexible case \cite{bernabei1,bulacu2,bernabei2,bernabei3}, 
and blending with the same A-homopolymer 
leads to a weaker dynamic asymmetry in the model II.

For times longer than $\tau_{\alpha}$, the MSD of both species 
exhibits subdiffusive behaviour over several time decades, 
prior to the final crossover to diffusive behaviour at
much longer times. The subdiffusive regime can be described 
by an effective power law $\langle \Delta r^2 (t)\rangle \sim t^{x}$,
with $x < 1$. This is a consequence of chain connectivity
and a characteristic feature of polymer systems.
In the case of fully-flexible homopolymers it reflects
Rouse dynamics. Figure~\ref{fig:msdasym}c shows the temperature 
dependence of the $x$-exponents for the two species
in both models I and II. Within statistics, the (fully-flexible) 
A-chains show the same temperature independent
exponent, $x \approx 0.64$, in both models. This is also the value
observed for the fully-flexible A- and B-homopolymers (not shown), and
can be easily understood in terms of Rouse dynamics. The value $x \approx 0.64$
is higher than the Rouse exponent $x = 0.5$, which is the effective value
predicted by the Rouse model in the limit $N \rightarrow \infty$ \cite{doibook}.
Thus, the former difference is mostly due to the finite size, $N=21$, of the chains \cite{bennemannrouse}.
We obtain a similar, almost $T$-independent, exponent $x \sim 0.64$ for the semiflexible
B-homopolymer. This coincidence is probably fortuitous, since  the $x$-value for the
semiflexible homopolymer cannot be assigned to Rouse dynamics, which does
not incorporate semiflexibility.

Concerning the exponents for the B-chains {\it in the blend}
they exhibit a rather different behaviour in the models I and II (see Figure~\ref{fig:msdasym}c). 
In the model I, the exponent $x$ decreases monotonically, taking values
much smaller than the $T$-independent value $x = 0.64$ found for the fully-flexible B-homopolymer. 
This trend reflects the breakdown of the Rouse model for the
fully-flexible B-chains in the blend, as discussed in \cite{prlrouse}. 
This breakdown is also reflected in the anomalous scaling observed for the Rouse modes
on decreasing $T$ (see below).
The lowest investigated temperature in the model II at which  $x$ can be solved is $T = 0.5$.
The dynamic asymmetry at this $T$ is the same that in the model I at $T \gtrsim 0.6$,
for which  $x$ has decreased by about a 15\% from its value at $T = 1.5$.
On the contrary, in the model II it remains, within statistics, 
constant with a value $x \sim 0.64$, as observed for the 
semiflexible B-homopolymer (see above).

\subsection{Rouse modes}

\begin{figure}
\begin{center}
\includegraphics[width=0.70\textwidth]{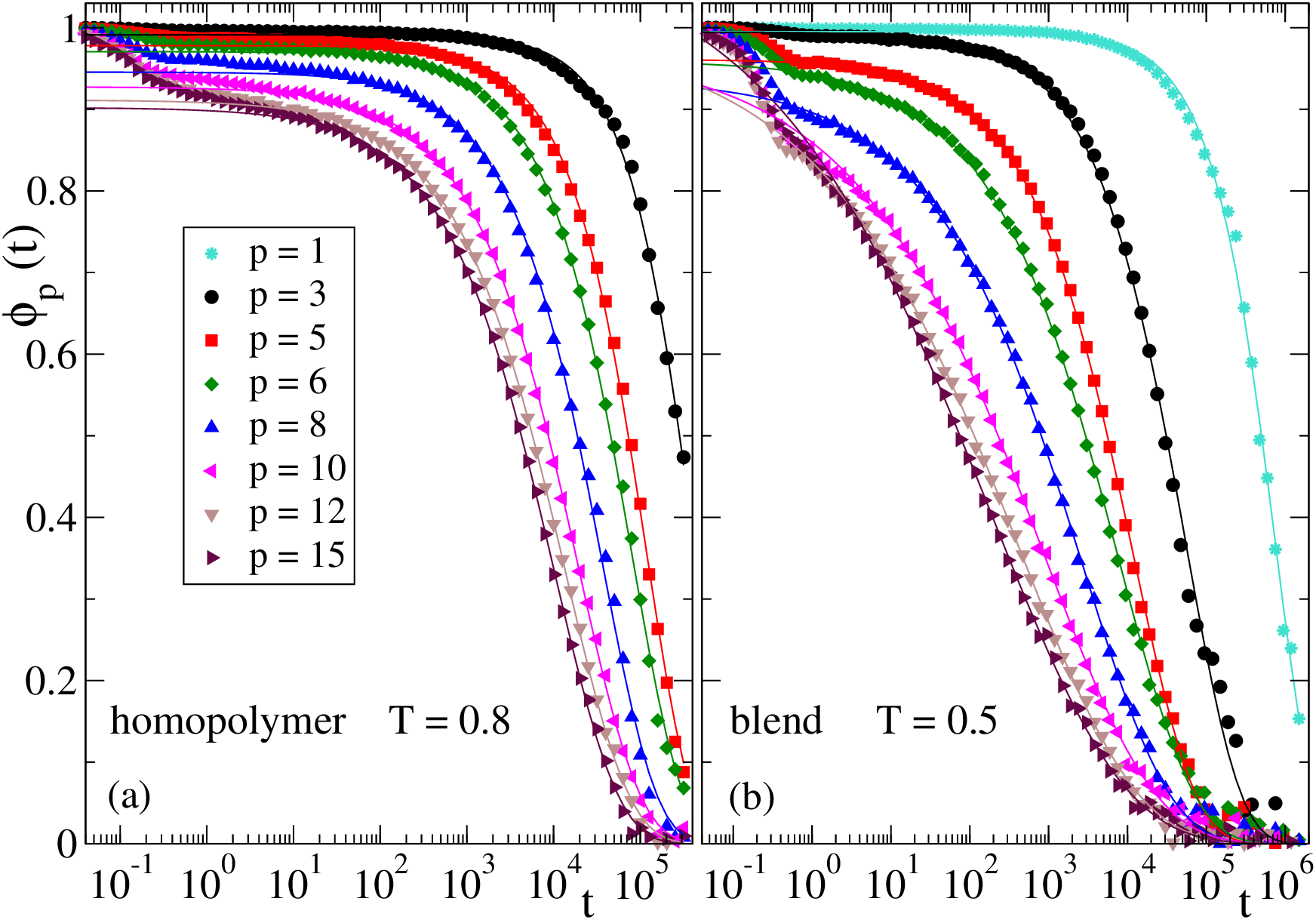}
\end{center}
\caption{Symbols: Rouse correlators for semiflexible B-chains in the homopolymer state (a)
and in the blend (b), at respectively $T = 0.8$ and $T = 0.5$. Lines
are fits to stretched exponentials (see text).}
\label{fig:rouse}
\end{figure}
\begin{figure}
\begin{center}
\includegraphics[width=0.85\textwidth]{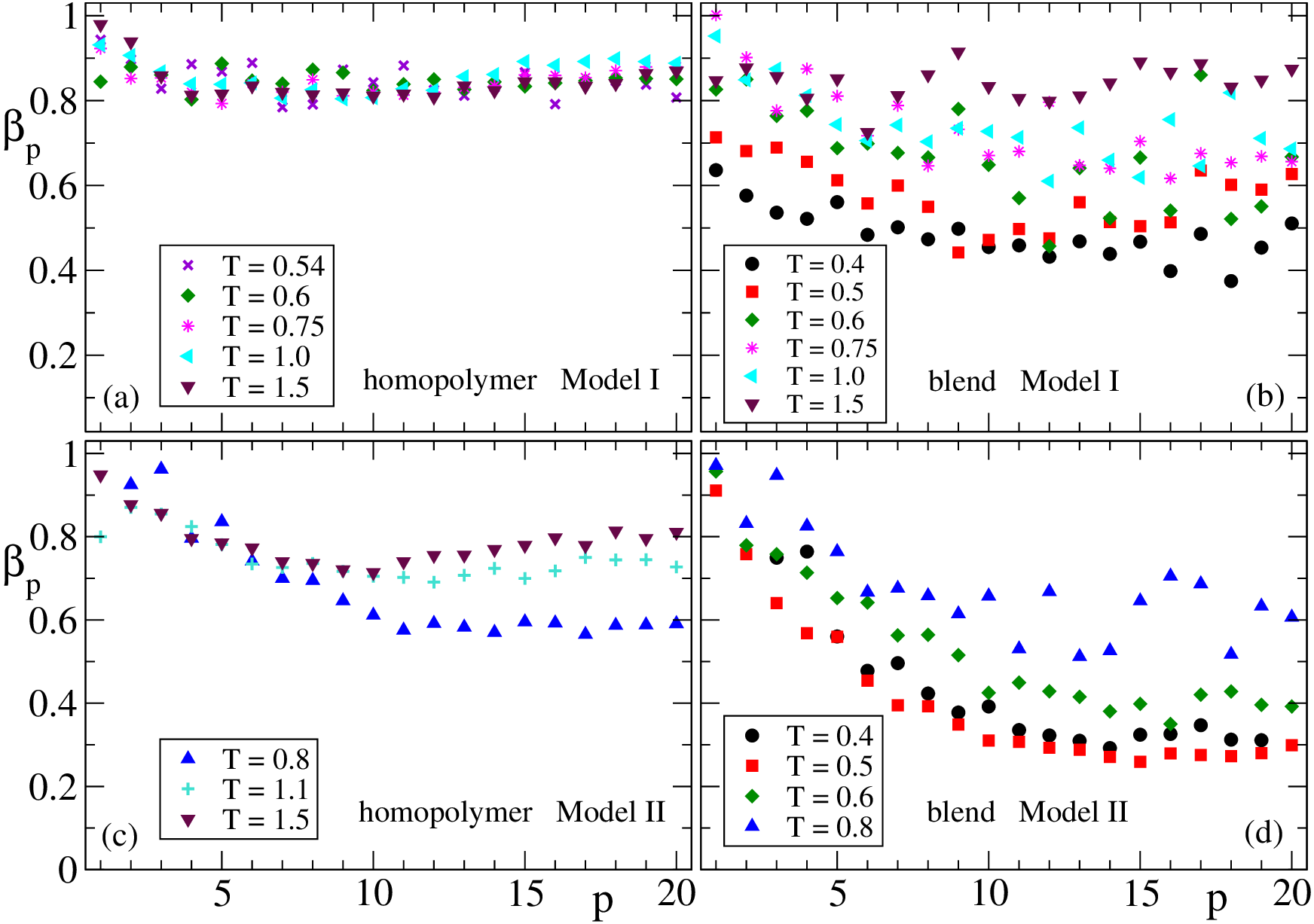}
\end{center}
\caption{$T$- and $p$-dependence of the stretching exponents of the Rouse correlators 
for the B-chains (see text). The data in the left panels correspond to the fully-flexible (a)
and semiflexible (c) B-homopolymer. The data in the right panels correspond to B-chains in the blend,
for the models I (b) and II (d).}
\label{fig:beta}
\end{figure}

The latter observations suggest that, unlike for fully-flexible polymers, scaling properties
for chain dynamics in semiflexible polymers may be unaltered by blending.
We confirm this point by analyzing the relaxation of the internal chain degrees of freedom.
This can be quantified by computing the correlators of the Rouse modes. 
The latter are defined as \cite{doibook}
\begin{equation}
{\bf X}_p (t) = N^{-1}\sum_{j=1}^{N}{\bf r}_j (t) \cos [(j-1/2)p\pi/N] ,
\label{eq:rouse}
\end{equation}
where ${\bf r}_j$ denotes the position of the $j$th-monomer of the tagged chain, 
and $0 \le p \le N-1$ is the mode index. The wavelength of each $p$-mode is given by $N/p$.
Therefore, smaller $p$-values probe relaxation at larger length scales.
The normalized Rouse correlators are defined as 
$\Phi_p (t) = \langle {\bf X}_p (t) \cdot {\bf X}_p (0) \rangle /\langle X^2_p (0) \rangle$.
According to the Rouse model \cite{doibook}, the correlators 
decay exponentially $\Phi_p (t) = \exp(-t/\tau_p)$,
and the relaxation times scale as $\tau_p \sim (N/p)^2$. This is also the scaling
behaviour of the static amplitudes of the modes, $\langle X^2_p (0) \rangle \sim (N/p)^2$,
and therefore $\tau_p \sim \langle X^2_p (0) \rangle$. 
Figure~\ref{fig:rouse} shows results of $\Phi_p (t)$ for semiflexible B-chains
in the homopolymer state and in the blend (model II) at two selected low temperatures. 
The first decay, at short times, to the plateau reflects 
the onset of the caging regime already observed in the MSD. The second decay at long times
reflects the relaxation of the corresponding $p$-mode.

\begin{figure}
\begin{center}
\includegraphics[width=0.75\textwidth]{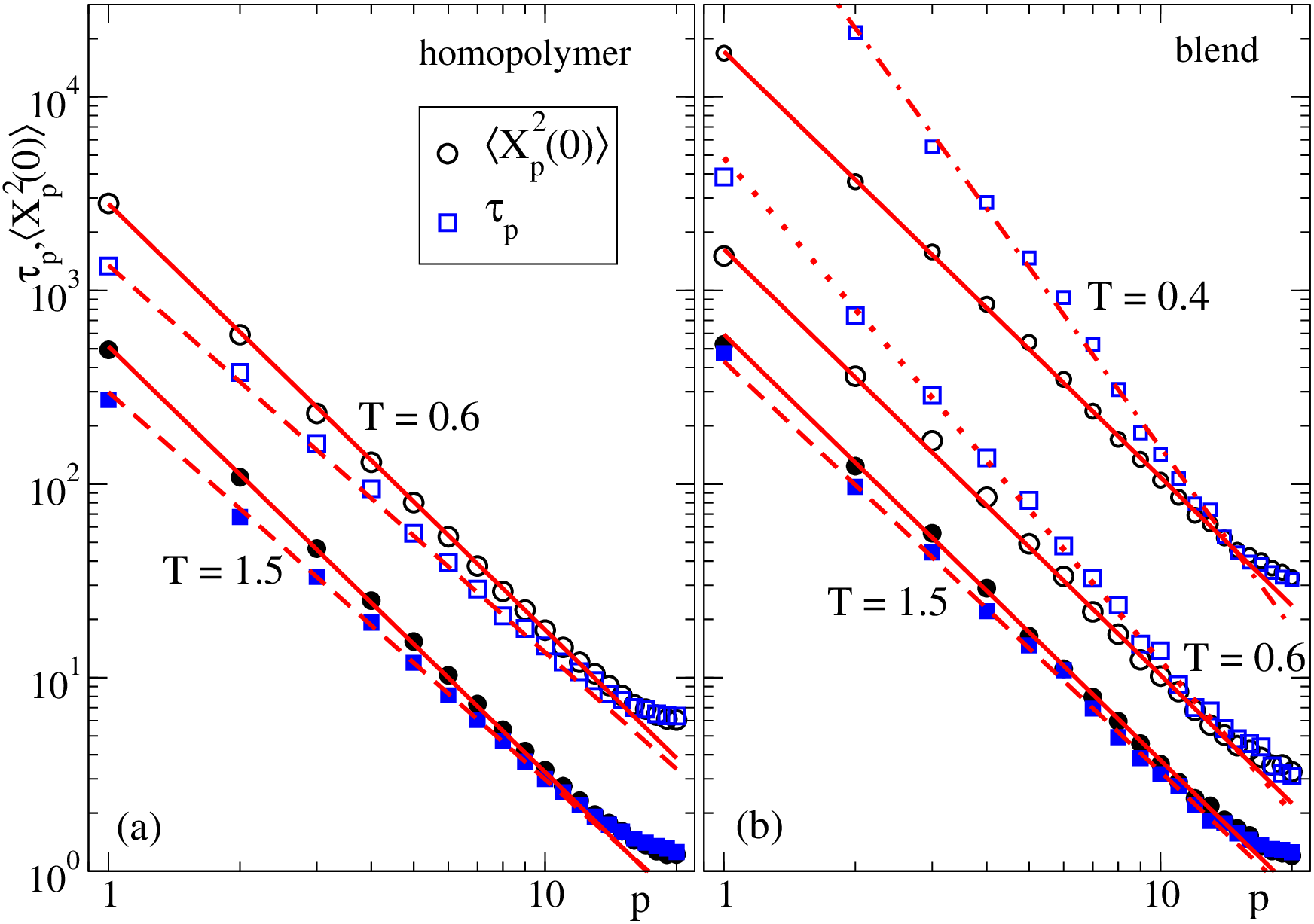}
\end{center}
\caption{Squares and circles are respectively the relaxation times $\tau_p$
and the static amplitudes $\langle X^2_p(0)\rangle$ of the Rouse correlators for the fully-flexible B-chains.
Panel (a): data for the fully-flexible B-homopolymer at $T =0.6$ (empty symbols) and $T = 1.5$ (filled symbols).
Panel (b): data for the fully-flexible B-chains in the blend, at $T = 0.4$ (small empty symbols),
$T =0.6$ (big empty symbols) and $T = 1.5$ (filled symbols). Units in the vertical axis are arbitrary.
Each data set has been rescaled by a factor to facilitate  comparison of $\tau_p$ and $\langle X^2_p(0)\rangle$ 
at common $T$. Solid and dashed lines in both panels
are power laws with respectively $x = 2.2$ and $x =2.0$. 
The dotted ($T = 0.6$) and dashed-dotted ($T = 0.4$) lines in panel (b) are power laws with 
respectively $x = 2.7$ and $x = 3.3$.}
\label{fig:tau-flex}
\end{figure}

We analyze the latter by fitting the decay to a stretched exponential 
or Kohlrausch-Williams-Watts (KWW) function 
$\Phi_p(t) = A_p \exp[-(t/\tau_p)^{\beta_p}]$, with $A_p, \beta_p < 1$.
The KWW time $\tau_p$ provides an estimation of the relaxation time of the $p$-mode.
Figure~\ref{fig:beta} shows, at several temperatures, 
the stretching exponents $\beta_p$ for the fully-flexible 
and semiflexible B-chains, both in the homopolymer and in the blend.
The general trend displayed by the four panels is that stretching is enhanced both by
the presence of intramolecular barriers and by blending.
The data for the fully-flexible B-homopolymer (Figure~\ref{fig:beta}a) 
are roughly $T$-independent and take values
close to Rouse-like exponential behaviour $\beta_p =1$ (see above). This is in agreement with observations
in similar fully-flexible models \cite{bennemannrouse}. A rather different behaviour is observed
for the fully-flexible B-chains in the blend (Figure~\ref{fig:beta}b). Exponential behaviour
is only approached at high temperatures, in the limit of vanishing dynamic asymmetry. 
On decreasing temperature and increasing the dynamic asymmetry, the correlators
exhibit stronger stretching, reaching values of even $\beta_p \sim 0.4$. 
It can be easily shown (see Ref. \cite{prlrouse}) that
if the KWW time shows Rouse scaling, $\tau_p \sim p^{-2}$,
stretching (i.e., $\beta_p < 1$) in $\Phi_p(t)$  esentially arises from a distribution of  
Rouse-like exponential processes (presumably related to dynamic heterogeneities). 
On the contrary, if $\tau_p$ 
strongly deviates from Rouse scaling,
the modes have an intrinsically strong non-exponential nature.
Figure~\ref{fig:tau-flex} shows results for the $p$-dependence of $\tau_p$ 
for the fully-flexible B-chains in the homopolymer and in the blend (Model I). 
The selected temperatures cover all the investigated range.
We also include the amplitudes of the modes $\langle X^2_p(0) \rangle$, 
and rescale the data sets by arbitrary factors
to facilitate comparison between  dynamic and static quantities 
at the different temperatures.
In agreement with similar fully-flexible models \cite{bennemannrouse},
data for the homopolymer are consistent with Rouse scaling,
$\tau_p \sim p^{-2}$, and therefore the Rouse modes are esentially exponential.
The observed dynamic Rouse scaling is consistent with the static scaling displayed 
by the amplitudes of the modes, $\langle X^2_p(0)\rangle \sim p^{-2.2}$, 
again in agreement with previous works \cite{bennemannrouse}
and very close to the ideal gaussian behaviour $\langle X^2_p(0)\rangle \sim p^{-2}$
expected within the Rouse model. Thus, Rouse relaxation times and amplitudes for the homopolymer
are roughly proportional, $\tau_p \sim \langle X^2_p(0)\rangle$.

The static scaling $\langle X^2_p(0)\rangle \sim p^{-2.2}$ observed in the fully-flexible B-homopolymer
is not altered by blending at any investigated temperature (Figure~\ref{fig:tau-flex}b).
On the contrary, dynamic Rouse scaling is observed in the blend only at high temperature.
On decreasing temperature and increasing the dynamic asymmetry, a progressive deviation
from the relation $\tau_p \sim \langle X^2_p(0)\rangle$ is observed.
We describe the behaviour of the relaxation times 
by an effective power law $\tau_p \sim p^{-x}$,
with $x$ increasing on decreasing temperature, up to $x = 3.3$ for $T = 0.4$.
Thus, dynamic asymmetry in the blend leads to an intrinsic strongly non-exponential 
character of the Rouse modes for the fully-flexible B-chains.
Intrinsic non-exponentiality and the observed anomalous scaling for the relaxation times 
are not related to particular
static features of the modes, which indeed are not affected by blending.
The origin of the stretching of the Rouse correlators for the semiflexible B-chains
[see panels (c) and (d) of Figure~\ref{fig:beta}] will be discussed below.

\begin{figure}
\begin{center}
\includegraphics[width=0.75\textwidth]{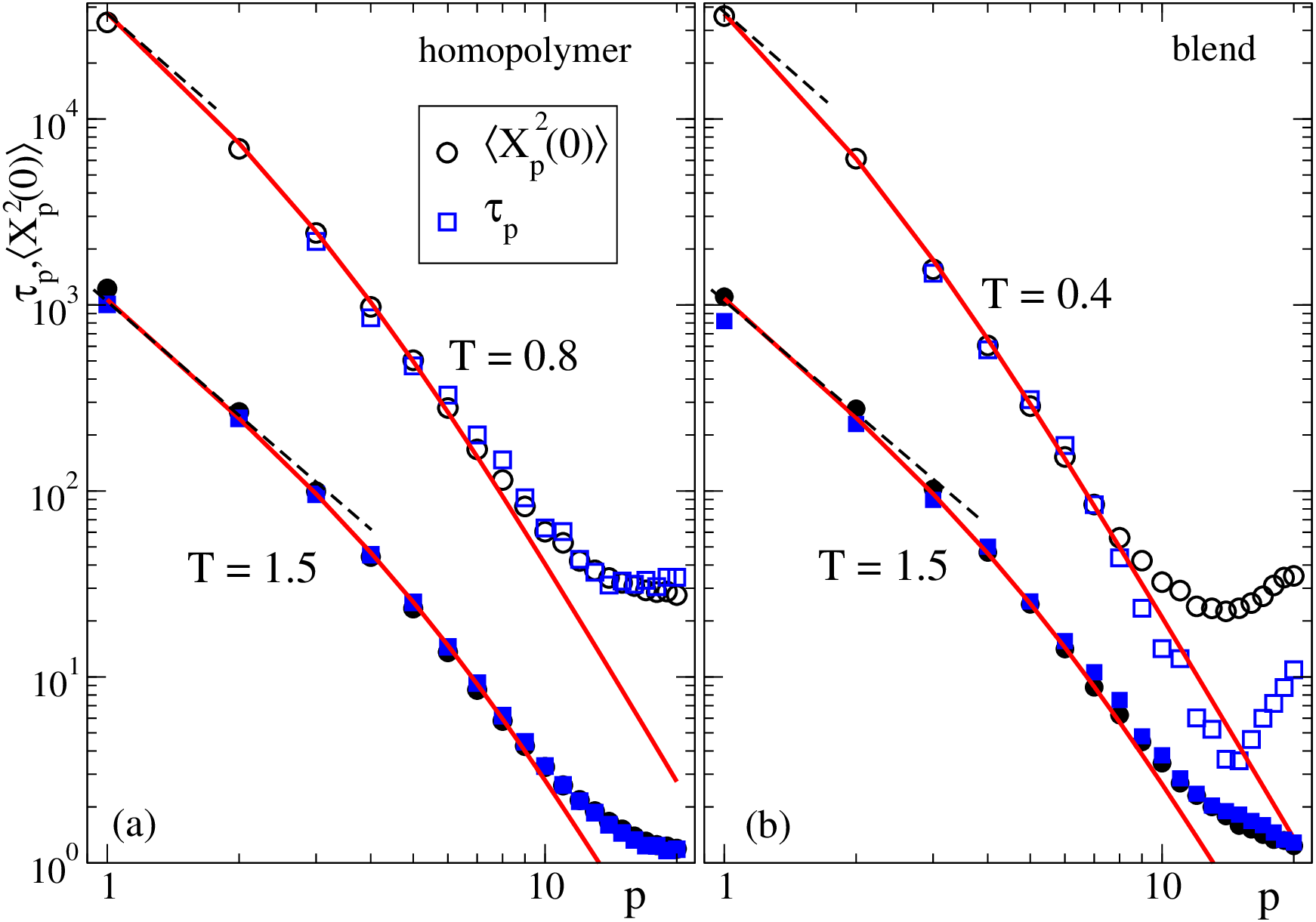}
\end{center}
\caption{Squares and circles are respectively the relaxation times $\tau_p$
and the static amplitudes $\langle X^2_p(0)\rangle$ of the Rouse correlators for the semiflexible B-chains.
Panel (a): data for the semiflexible B-homopolymer at $T =0.8$ (empty symbols) and $T = 1.5$ (filled symbols).
Panel (b): data for the semiflexible B-chains in the blend, at 
$T =0.4$ (empty symbols) and $T = 1.5$ (filled symbols). Units in the vertical axis are arbitrary.
Each data set has been rescaled by a factor to facilitate  comparison of $\tau_p$ and $\langle X^2_p(0)\rangle$ 
at common $T$. Solid lines in both panels
are fits to the Equation~(\ref{eq:tau-stiff}).
Dashed lines represent Rouse-like behaviour $\tau_p , \langle X^2_p(0)\rangle \sim p^{-2}$.}
\label{fig:tau-stiff}
\end{figure}

Figure~\ref{fig:tau-stiff} shows results of $\tau_p$ and $\langle X^2_p (0)\rangle$
for the semiflexible B-chains in the homopolymer and in the blend (Model II), at temperatures
covering all the investigated range. In analogy with the representation of Figure~\ref{fig:tau-flex},
we rescale the different data sets to facilitate comparison between times and amplitudes.
This procedure yields a nice overlap of the latter for the B-homopolymer, and as in the
fully-flexible case, the data follow the approximate relation $\tau_p \sim \langle X^2_p(0)\rangle$.
This is also the case for the semiflexible B-chains in the blend at high $T$.
The relation $\tau_p \sim \langle X^2_p(0)\rangle$
is also maintained at $T = 0.4$ except for modes $p > 9$, corresponding to small wavelengths $N/p \lesssim 2.3$.
However, it is noteworthy that the static data of Figure~\ref{fig:tau-stiff} 
follow a rather different $p$-dependence
from the gaussian behaviour $\langle X_p^2(0)\rangle \sim p^{-2}$, which
is only approached in the limit $p \rightarrow 1$ at high $T$. This confirms the non-gaussian character
of the semiflexible chains within all their length scale (see Introduction).
Obviously, also the $p$-dependence of the times is very different from the scaling
$\tau_p \sim p^{-2}$ predicted by the Rouse model.

We discuss the behaviour of $\tau_p$ and $\langle X_p^{2} (0)\rangle$
in terms of the framework proposed by Harnau and co-workers \cite{harnau1,harnau2,harnau3}.
In this approach, the equations of motion for the tagged chain are derived from
the Hamiltonian of the worm-like chain model \cite{rubinstein}. This leads to modified
Rouse equations of motion, which include intrachain bending forces not present in the
original Rouse model for fully-flexible chains. As in the original Rouse model, the interactions
of the tagged chain with the surroundings are simply modelled in terms of a friction term
and random forces \cite{harnau1,harnau2,harnau3}. Solution of the equations of motion
lead to a modified $p$-dependence of the amplitudes and relaxation times of the Rouse modes,
but as in the original Rouse model, $\langle X_p^{2} (0)\rangle$ and $\tau_p$ are proportional. 
The obtained relation is  
\begin{equation}
\tau_p \sim \langle X_p^{2} (0)\rangle \sim \left[p^2 +\frac{\pi^2\l^2}{N^2\langle b\rangle^2}p^4 \right]^{-1},
\label{eq:tau-stiff}
\end{equation}
where $\l$ is the persistence length of the chain. Recent simulation results
by Steinhauser and co-workers \cite{steinhauser,steinhauserrev} 
on semiflexible homopolymers are in good agreement with Equation~(\ref{eq:tau-stiff}). 
On this basis, we use it for describing the data sets of Figure~\ref{fig:tau-stiff}.
Instead of using different definitions of the persistence length proposed
in the literature (see e.g., Ref.~\cite{hsu}) as input, we just obtain $\l$  
as a fit parameter. This is forced to be identical for $\langle X_p^{2} (0)\rangle$ and $\tau_p$
at a same temperature, to be consistent with the relation $\tau_p \sim \langle X^2_p(0)\rangle$. 
The obtained values of the persistence length change from $\l \sim 1.2$ at high temperatures
to $\l \sim 3$ at the lowest investigated $T$ in the blend.
Equation (6) provides a good description of the data of Figure~\ref{fig:tau-stiff} 
in the region $p < 9$, both for the B-homopolymer and for the B-chains in the blend. 
The behaviour at smaller wavelengths cannot be captured. Presumably
this is mostly due to the influence of the torsional terms, which are not accounted for within
the approach of Refs.~\cite{harnau1,harnau2,harnau3}.  

\section{Discussion}

As shown in the previous section, the results presented in Figures~\ref{fig:tau-flex} 
and \ref{fig:tau-stiff} for the B-homopolymers are consistent with the approximate relation
$\tau_p \sim \langle X_p^2(0) \rangle$, the specific $p$-dependence being distinct
for fully-flexible and semiflexible chains. The latter is Rouse-like for fully-flexible
chains, $\tau_p \sim p^{-2}$, and is well described by Equation~(\ref{eq:tau-stiff})
for semiflexible chains. Since the relation $\tau_p \sim \langle X_p^2(0) \rangle$ is fulfilled,
the anomalous (in the meaning of non-Rouse) scaling of the relaxation times for the
semiflexible homopolymers is esentially of static origin. More specifically, it is
a direct consequence of the non-gaussian nature of the static intramolecular correlations.
This is also the case for the semiflexible B-chains in the blend (Model II), for which
$\tau_p \sim \langle X_p^2(0) \rangle$ is maintained,
except for short wavelengths ($N/p \lesssim 2.3$, see Figure~\ref{fig:tau-stiff}b)
at the lowest investigated temperatures. 

According to the approach 
of Refs.~\cite{harnau1,harnau2,harnau3} for semiflexible chains, the Rouse correlators $\Phi_p (t)$
decay, as in the Rouse model, exponentially. Following the same argumentation
as for the fully-flexible case (see Section 3), stretching in $\Phi_p (t)$
arises from a distribution of the predicted exponential processes if the KWW times 
fulfill the relation $\tau_p \sim \langle X_p^2(0) \rangle$. Otherwise
non-exponentiality is intrinsic. Thus, from data in Figure~\ref{fig:tau-stiff},
we conclude that non-exponentiality of the Rouse modes
of the semiflexible chains is intrinsinc only for short wavelengths,
in the blend state and at the lowest investigated temperatures.

As shown by the data in Figure~\ref{fig:msdasym}b, the dynamic asymmetry ($\chi \approx 68$) 
at the lowest investigated $T = 0.4$ of the Model II, is slightly higher
than that of the Model I at $T =0.6$ ($\chi \approx 63$). Though the strength of the confinement
is esentially the same in both cases, the origin of anomalous chain dynamics seems to be very
different. As discussed above, this is of intramolecular and static nature for the
semiflexible B-chains of Model II. On the contrary, the origin is esentially dynamic
for the fully-flexible B-chains of Model I, as indicated by the clear
breakdown of the relation $\tau_p \sim \langle X_p^2(0)\rangle$. As shown in Figure~\ref{fig:tau-flex}b,
we find strongly non-Rouse scaling for the times, $\tau_p \sim p^{-x}$ with $x = 2.7$,
considerably {\it larger} than the exponent $x = 2.2$, close to gaussian behaviour, found for the amplitudes.
Note that for the homopolymer we find a, Rouse-like, 
dynamic exponent $x = 2.0$ similar but {\it smaller} than the static $x =2.2$ (Figure~\ref{fig:tau-flex}a).

As mentioned in the Introduction, within the Rouse model memory effects, related to slow density fluctuations
of the matrix around the tagged chain, are neglected. The interactions of the tagged (gaussian) chain
with the surroundings are simply modelled by a friction term and random forces \cite{doibook}.
This Markovian approximation is also followed by the approach of Harnau and co-workers
~\cite{harnau1,harnau2,harnau3}, which just incorporates bending forces in the Rouse equations
of motion to account for chain stiffness. The predicted scaling behaviour
[Equation (\ref{eq:tau-stiff})] is observed for the semiflexible B-homopolymer
and is not signifficantly affected by blending with a slower matrix, suggesting 
that memory effects are not relevant and the Markovian approximation can still be applied in the blend.
It remains to be understood, from a microscopic basis,
why memory effects induced by the matrix are apparently much weaker
than intramolecular effects induced by the presence of the barriers. 
On the contrary, memory effects are crucial in the case of the fully-flexible 
B-chains in the blend \cite{prlrouse},
for which predictions of the Rouse model are strongly violated.

Finally, we want to make a last remark on the dynamics of the semiflexible chains in the blend. 
As discussed above, the semiflexible chains considered here are strongly non-gaussian
within all chain length scales. If longer chains were considered, with the same torsional
and bending contributions, the semiflexible character would be lost beyond some mode wavelength, 
and gaussian statistics would be recovered for intrachain static correlations at large length scales.
We expect that in such length scales, as observed for the fast fully-flexible 
chains in blends with strong dynamic asymmetry,
anomalous dynamic scaling will arise for the Rouse modes, distinct from the static gaussian scaling,
and memory effects will be relevant.
In fact this seems to be the case for the results reported in Ref.~\cite{brodeck} 
from atomistic simulations of the dynamically asymmetric blend PEO/PMMA.
There the relaxation times $\tau_p$ of the Rouse mode correlators of the fast component (PEO)
still exhibit anomalous, non-Rouse, $p$-scaling at large mode wavelengths $N/p$ 
for which gaussian behaviour, $\langle X^2_p (0)\rangle \sim (N/p)^2$, is recovered in the mode amplitudes
(see Figure 11 in Ref.~\cite{brodeck}).
This would be the general expected behaviour in real blends with strong dynamic asymmetry.

\section{Conclusions}

We have performed simulations of a simple model of non-entangled semiflexible chains
blended with a slower component. Extending previous investigations for fully-flexible chains,
we investigate the effect of the dynamic asymmetry in the blend on the relaxation
of the semiflexible chains. We find the same anomalous scaling behaviour for the relaxation times
and the static amplitudes of the Rouse modes, in agreement with Markovian models
which extend the Rouse equations by the introduction of bending forces. 
Thus, anomalous dynamic features for the semiflexible chains  
esentially have a static and intramolecular origin.
This is very different from the case of fully-flexible chains in blends with similar dynamic asymmetry. 
For the latter anomalous dynamic scaling is strongly correlated with the dynamic asymmetry,
and not to  features of the static amplitudes, which indeed still follow gaussian scaling.

The former results suggest that  memory effects
induced by the surrounding slow matrix are not relevant for non-entangled semiflexible polymers,
and Markovian models can still be applied. However, if the chains are long enough,
we expect that the influence of the memory effects will be recovered at large length scales
where intrachain static correlations recover gaussian statistics.  
Work in this direction is in progress.

\section{Acknowledgments}

We acknowledge financial support from the projects FP7-PEOPLE-2007-1-1-ITN (DYNACOP, EU),
MAT2007-63681 (Spain), and IT-436-07 (GV, Spain).

\end{document}